\RequirePackage{fix-cm}
\documentclass[smallextended]{svjour3}       
\smartqed  
\usepackage{graphicx}
\usepackage{amsmath, amssymb, amsfonts, bm}
\usepackage{hyperref}
\usepackage{xcolor}
\begin{document}

\title{Multilevel effects in a driven generalized Rabi model
}


\author{I.~Pietik\"ainen \and
S.~Danilin \and
K.~S. Kumar \and
J.~Tuorila \and
G.~S.~Paraoanu
}


\institute{I. Pietik\"ainen \at Nano and Molecular Systems Research Unit, University of Oulu, P.O. Box 3000, FI-90014, Finland\\
\email{iivari.pietikainen@oulu.fi} \and
S.~Danilin, K.~S.~Kumar, G.~S.~Paraoanu \at
Centre for Quantum Engineering and LTQ, Department of Applied Physics, Aalto University, P.O. Box 15100, FI-00076 Aalto, Finland \\
\email{sorin.paraoanu@aalto.fi} \and
J. Tuorila \at COMP Centre of Excellence, Department of Applied Physics, Aalto University, P.O. Box 15100, FI-00076 Aalto, Finland \\
Nano and Molecular Systems Research Unit, University of Oulu, P.O. Box 3000, FI-90014, Finland
}

\date{Received: date / Accepted: date}

\maketitle

\begin{abstract}
We study numerically the onset of higher-level excitations and resonance frequency shifts in the generalized multilevel Rabi model with dispersive coupling under strong driving. The response to a weak probe is calculated using the Floquet method, which allows us to calculate the probe spectrum and extract the resonance frequency. We test our predictions using a superconducting circuit consisting of transmon coupled capacitively to a coplanar waveguide resonator. This system is monitored by a weak probe field, and at the same time driven at various powers by a stronger microwave tone. We show that the  transition from the quantum to the classical regime is accompanied by a rapid increase of the transmon occupation and, consequently that the qubit approximation is valid only in the extreme quantum limit.

\end{abstract}

\section{Introduction}
\label{intro}

The interaction between light and matter has been traditionally studied in terms of the Jaynes-Cummings (JC) model~\cite{Jaynes63}. In the quantum optical nomenclature, the JC Hamiltonian describes an atomic transition linearly coupled to an electromagnetic field mode. In optical systems the coupling is typically weak, but by placing the atom inside a Fabry-Perot cavity, it can be increased by decreasing the mode volume of the cavity~\cite{Devoret07}. 
Classically, the interaction is reminiscent of that between two harmonic oscillators, and the system resembles two independent oscillators with natural frequencies given by the normal modes~\cite{Zhu90}. In the dispersive regime, the coupling of the oscillators is much smaller than the detuning between their natural frequencies and, consequently, one normal-mode frequency is close to the natural frequency of the cavity and the other to the transition frequency of the atom. The quantum nature of the system can be put in evidence by observing the non-equidistant transition frequencies in the spectrum caused by the inherently anharmonic atom~\cite{Carmichael96,Fink08}.

Superconducting circuits have performed excellently in reproducing quantum optical phenomena observed in conventional cavity-quantum electrodynamical (cQED) systems~\cite{Fink08,Blais04,Wallraff04,Fragner08,Silveri15,Li13}. This experimental platform allows the manufacturig of complex superconducting circuits with controlled design parameters \cite{Schoelkopf08,Paraoanu14}. 
The scalability of this architecture is an advantage for the potential use in the quantum information devices. A paradigmatic circuit analogue of the cQED setup is obtained by coupling a transmon qubit with a coplanar waveguide resonator~\cite{Koch07,Schuster07}, where the transmon represents the anharmonic atomic degree of freedom and the resonator is the linear cavity element. 

The quantum Rabi model refines the JC Hamiltonian by including terms that do not conserve the occupation number. The corrections imposed by these terms become observable in the ultra-strong coupling limit, where even the ground state of the coupled system has non-zero occupations of its constituent parts~\cite{Ciuti06,Lolli15,Cirio17}. This is reflected also in the weak probe spectrum as the vacuum Bloch-Siegert shift of the spectral lines~\cite{Bloch40,FornDiaz10} from their values given by the normal modes of the JC model. The total deviation between the spectral lines and the bare resonance frequencies of the constituent parts is called the vacuum Stark shift. By driving the system, while at the same time applying a weak probe tone to the cavity, the resulting spectrum is expected to show multiple lines displaying the nonlinear level structure of the system, as the cavity is excited by the pump. At high powers of the drive, the qubit eventually saturates~\cite{Alsing92}, even in the case of a dispersive coupling. Consequently, the spectral lines converge with increasing power into peaks centered at the bare cavity and qubit transition frequencies~\cite{Bishop09}. This continuous transition between essentially quantum and purely classical responses can be interpreted as the breakdown of the photon blockade of the cavity~\cite{Carmichael15,Fink17}, and is also referred to as the driven quantum-to-classical transition~\cite{Pietikainen17}. Instead of driving, the transition can be observed also by increasing the temperature of the system~\cite{Fink10}.

The power dependence of the spectral lines is referred to as the dynamic Stark shift~\cite{Silveri17}. The dynamic Stark shift has been studied experimentally with superconducting circuits in Refs.~\cite{Tuorila10,Tuorila13}, where the qubit instead of the cavity was driven. The strongly driven dispersive regime has been used to read out the qubit state, which is displayed in the transient dynamics~\cite{Reed10,Bishop10,Boissonneault10}. However, because the transmon is only weakly anharmonic, the strong driving of the cavity excites it from the low-energy subspace defined by the two lowest energy eigenstates. In such case, the model has to be improved by including the higher eigenstates of the transmon~\cite{Boissonneault12a,Boissonneault12b,Pietikainen17}. Such system can be discussed in terms of a so-called generalized Rabi model.


In this paper, we numerically study the transmon state occupation as the drive power applied to the cavity is swept across the quantum-to-classical transition.  The numerical simulations were done by calculating the probe reflection coefficient using the Floquet method~\cite{Grifoni98,Silveri13}.
We compare the experimental data with the simulations with different truncations of the transmon state space. We demonstrate that the truncation of the transmon to two states (qubit approximation) becomes inaccurate already early in the transition between the quantum and the classical phases, and we study the convergence of the simulation for up to five transmon states.



\section{Driven multilevel quantum Rabi model}
\label{Rabi}

\begin{figure*}
  \includegraphics[width=1\textwidth]{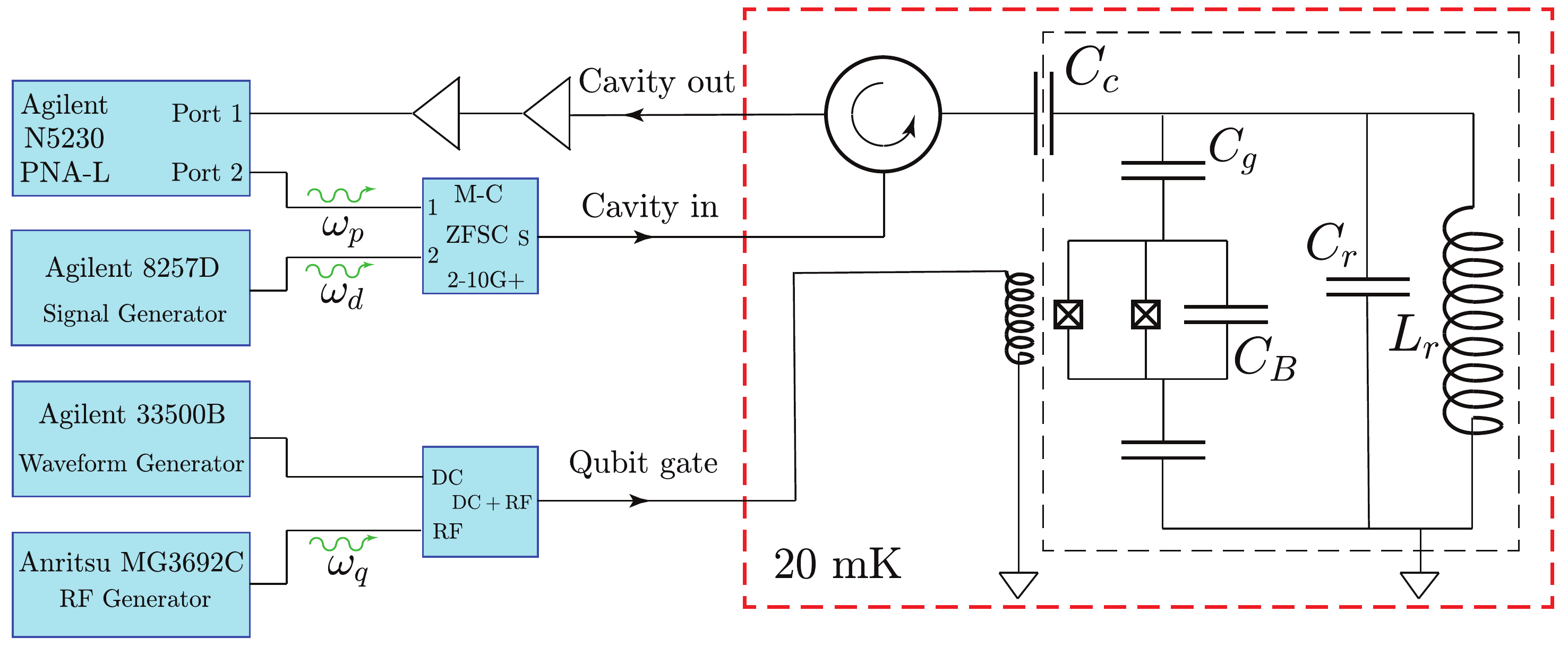}
\caption{Circuit schematic of the experimental setup. The generalized Rabi system is indicated in the figure with a black dashed box. }
\label{fig:schema}
\end{figure*}

In Fig. \ref{fig:schema}, we show a simplified circuit schematic of our experimental setup. The measured system consists of a $\lambda/4$-waveguide-resonator capacitively coupled to a transmon. The control electronics are operated at the room temperature while the actual circuit is placed inside a dilution refrigerator at a low temperature (20 mK). The resonator is modeled as an $LC$-oscillator with the bare resonance of frequency $\omega_{\rm c}/(2\pi) = 4.114$ GHz and a dissipation rate of $\kappa/(2\pi) = 4$ MHz. The energy-level structure of the transmon can be controlled by the flux gate and for this measurement the transition frequency of the lowest transition was set to $\omega_0/(2\pi) = 4.845$ GHz. We performed pump-probe measurements where the tone with frequency $\omega_{\rm d}$ was used for driving the system while another weaker tone with frequency $\omega_{\rm p}$ was used for probing the resonance frequency of the system. The probe frequency was sweeped across the bare cavity frequency at different drive powers and the reflection spectrum of the system was recorded with a spectrum analyzer. This method is the same as in Ref.~\cite{Pietikainen17} but the sample used in the experiments below is different.

The Hamiltonian for the driven cavity-transmon system depicted in Fig.~\ref{fig:schema} can be written into the form of a driven generalized Rabi Hamiltonian (scaled with $\hbar$) with $N$ transmon states
\begin{equation}\label{eq:ManyStatesHam}
\hat{H}= \omega_{\rm c} \hat{a}^{\dag}\hat{a} + \sum_{\sigma=0}^{N-1} \Omega_{\sigma} \hat{\Pi}_{\sigma\sigma} + (\hat{a}^{\dag}+\hat{a})\sum_{\sigma,\sigma'=0}^{N-1} g_{\sigma\sigma'}\hat{\Pi}_{\sigma\sigma'}+A\cos(\omega_{\rm d} t)(\hat{a}^{\dag}+\hat{a}),
\end{equation}
where $\hat{a}$ and $\hat\Pi_{\sigma\sigma'}\equiv |\sigma\rangle\langle \sigma'|$ are the ladder operators for the cavity and the transmon, and $|\sigma\rangle$ are the eigenstates of the transmon, $\sigma=0,1,2...$ We have also denoted the cavity and drive frequencies with $\omega_{\rm c}$ and $\omega_{\rm d}$, respectively. The eigenfrequency corresponding to the state $|\sigma\rangle$ is given by $\Omega_\sigma$, and $g_{\sigma\sigma'}$ describe the strength of a transition $\sigma'\rightarrow \sigma$ accompanied with an exchange of cavity photon. The amplitude of the drive is set by $A$.

For the numerical solution, it is beneficial to transform the system into the displaced frame using the time-dependent unitary transformation
\begin{equation}
\hat{D}(\alpha) = e^{\alpha\hat{a}^\dagger -\alpha^*\hat{a}}.
\end{equation}
By assuming that $\alpha$ obeys the classical equation of motion
\begin{equation}
\dot{\alpha} = -i\omega_{\rm c}\alpha - i \frac{A}{2}e^{-i\omega_{\rm d}t} - \frac{\kappa}{2}\alpha,
\end{equation}
one can write the master equation for the density operator of the coupled system into the form
\begin{equation}
\frac{d\hat{\rho}}{dt}= - i [\hat{H}_{\rm eff},\hat{\rho}] + \kappa \mathcal{L}[\hat{a}]\hat{\rho},
\end{equation}
where we work in the bad-cavity limit where $\kappa$ is much larger than the decay rates of the transmon. In the above, we have assumed that we operate in the low-temperature regime, and we defined the Lindblad superoperator $\mathcal{L}[\hat{A}]\hat{\rho} = \frac12 (\hat{A}\hat{\rho} \hat{A}^{\dag} - \hat{A}^{\dag}\hat{A}\hat{\rho}-\hat{\rho} \hat{A}^{\dag}\hat{A})$ and the effective Hamiltonian
\begin{eqnarray}
\label{eq:multiHam}
\hat{H}_{\rm eff}=&& \omega_{\rm c} \hat{a}^{\dag}\hat{a} + \sum_{\sigma=0}^{N-1} \Omega_{\sigma} \hat\Pi_{\sigma\sigma} + \left[e^{i\omega_{\rm d} t} + e^{-i\omega_{\rm d} t}\right]\sum_{\sigma,\sigma'=0}^{N-1} G_{\sigma\sigma'}\hat\Pi_{\sigma\sigma'} \nonumber\\
&&+(\hat{a}^\dagger +\hat{a})\sum_{\sigma,\sigma'=0}^{N-1} g_{\sigma\sigma'}\hat\Pi_{\sigma\sigma'},
\end{eqnarray}
where $G_{\sigma\sigma'} = g_{\sigma\sigma'}|\alpha_{\rm ss}|$ and 
\begin{equation}\label{eq:ass}
\alpha_{\rm ss} =\frac{Ae^{-i\omega_{\rm d} t}}{2\sqrt{\delta^2+\kappa^2/4}}
\end{equation}
is the steady state solution for $\alpha$, where $\delta = \omega_{\rm c}-\omega_{\rm d}$. We have also assumed a constant phase shift $\theta = \arctan[\kappa/(2\delta)]$ for the drive. We thus see that after the transformation the cavity is displaced into the vacuum and, as a consequence, the drive acts on the transmon instead. This is preferable in terms of a numerical solution as one expects that strong drive does not excite the detuned and anharmonic transmon as much as the nearly resonant and harmonic cavity, thus enabling truncation to a smaller set of basis states. 

We can form an intuitive picture by considering only the two lowest states of the transmon ($N=2$). By employing the rotating wave approximation and the Schrieffer-Wolff transformation on Eq. (\ref{eq:multiHam}), we obtain the driven Jaynes-Cummings Hamiltonian
\begin{equation}
\hat{H} = (\omega_{\rm c} +\chi_0\hat{\sigma}_{\rm z})\hat{a}^\dagger\hat{a} +\frac{1}{2}(\omega_0 +\chi_0)\hat{\sigma}_{\rm z} +\frac{G}{2}(e^{i\omega_{\rm d}t}\hat{\sigma}_- +e^{-i\omega_{\rm d}t}\hat{\sigma}_+),
\end{equation}
where $\omega_0 = \Omega_1-\Omega_0$, $g=g_{01}=g_{10}$ and $\chi_0 = g^2/(\omega_0-\omega_{\rm c})$ is the vacuum Stark shift in the rotating wave approximation. At low drive powers, the qubit is in ground state and the effective cavity frequency is shifted to a lower frequency $\omega_{\rm c}-\chi_0$. At high powers, the qubit becomes saturated and the dynamic Stark shift becomes averaged out resulting in an effective cavity frequency equal to the bare cavity frequency $\omega_{\rm c}$. 

We emphasize that the above simplified discussion neglects the influence of the higher transmon states on the Stark shift. Instead of solving the multilevel corrections perturbatively~\cite{Boissonneault12a,Boissonneault12b}, we solve numerically the Floquet eigenproblem of the $\tau=2\pi/\omega_{\rm d}$-periodic Hamiltonian~(\ref{eq:multiHam}). We form the Floquet matrix for the Hamiltonian in Eq.~(\ref{eq:multiHam}) in the basis defined by $\{\vert\sigma, n,l\rangle = \vert\sigma\rangle \otimes \vert n\rangle \otimes \vert l\rangle\}$, where $\vert\sigma\rangle$, $\sigma=0,1,...,N-1$ are the transmon states, $\vert n \rangle$, $n\in\mathbb{N}$ are the cavity states and $\vert l \rangle$, $l\in\mathbb{Z}$ denote the basis for $\tau$-periodic functions~\cite{Silveri13}. To obtain the numerical solution, one has to truncate also the bases of the cavity and $\tau$-periodic functions. We denote the maximum number of states in these bases with $N_{\rm c}^{\rm max}$ and $N_{\tau}^{\rm max}$, respectively. In the following, we refer to $N$, $N_{\rm c}^{\rm max}$ and $N_{\tau}^{\rm max}$ as the truncation numbers. By numerically solving the quasienergies and the corresponding quasienergy states of the Floquet matrix we can calculate the probe reflection coefficient $\Gamma(\omega_{\rm p}) = (Z(\omega_{\rm p})-Z_0)/(Z(\omega_{\rm p})+Z_0)$, where $Z(\omega_{\rm p})$ is the impedance of the driven system evaluated at the probe frequency $\omega_{\rm p}$, and $Z_0$ is the impedance of the transmission line which is used to propagate the measurement signal. The impedance $Z(\omega_{\rm p})$ is found by calculating the probe-induced transition rates and then employing the Kramers-Kronig relation~\cite{Tuorila13,Pietikainen17,Silveri13,LLSP}. More detailed explanation of the simulations can be found in the supplementary material of Ref.~\cite{Pietikainen17}.


\section{Experimental and numerical results}
\label{results}

\begin{figure*}[h!]
  \includegraphics[width=1\textwidth]{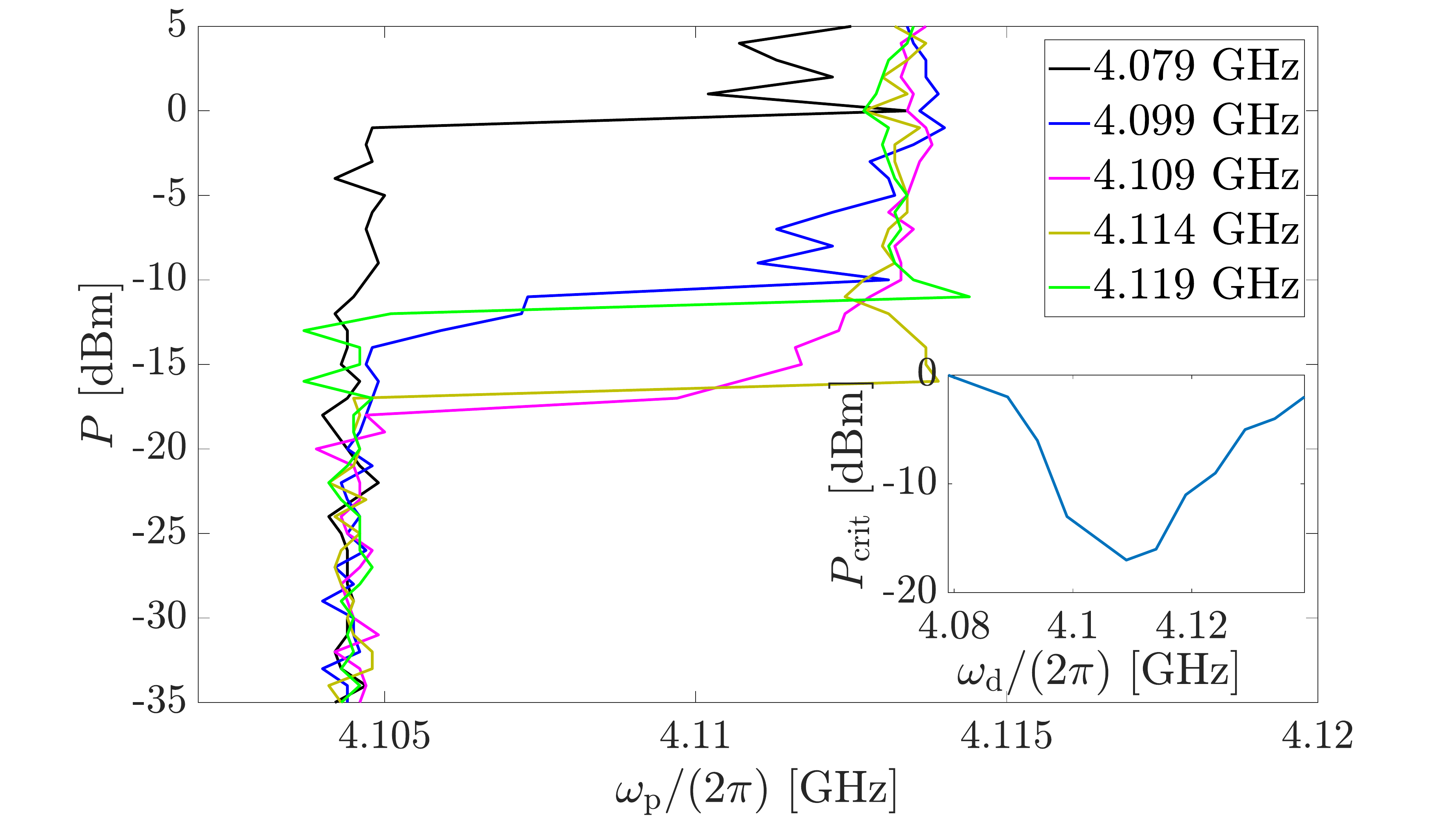}
\caption{The experimental resonance frequency of the system as a function of probe frequency $\omega_{\rm p}$ and drive power $P$ at different drive frequencies $\omega_{\rm d}$. Inset: The critical power $P_{\rm crit}$ as a function of the drive frequency $\omega_{\rm d}$.}
\label{fig:Experiments}
\end{figure*}

\begin{figure}
  \includegraphics[width=1\textwidth]{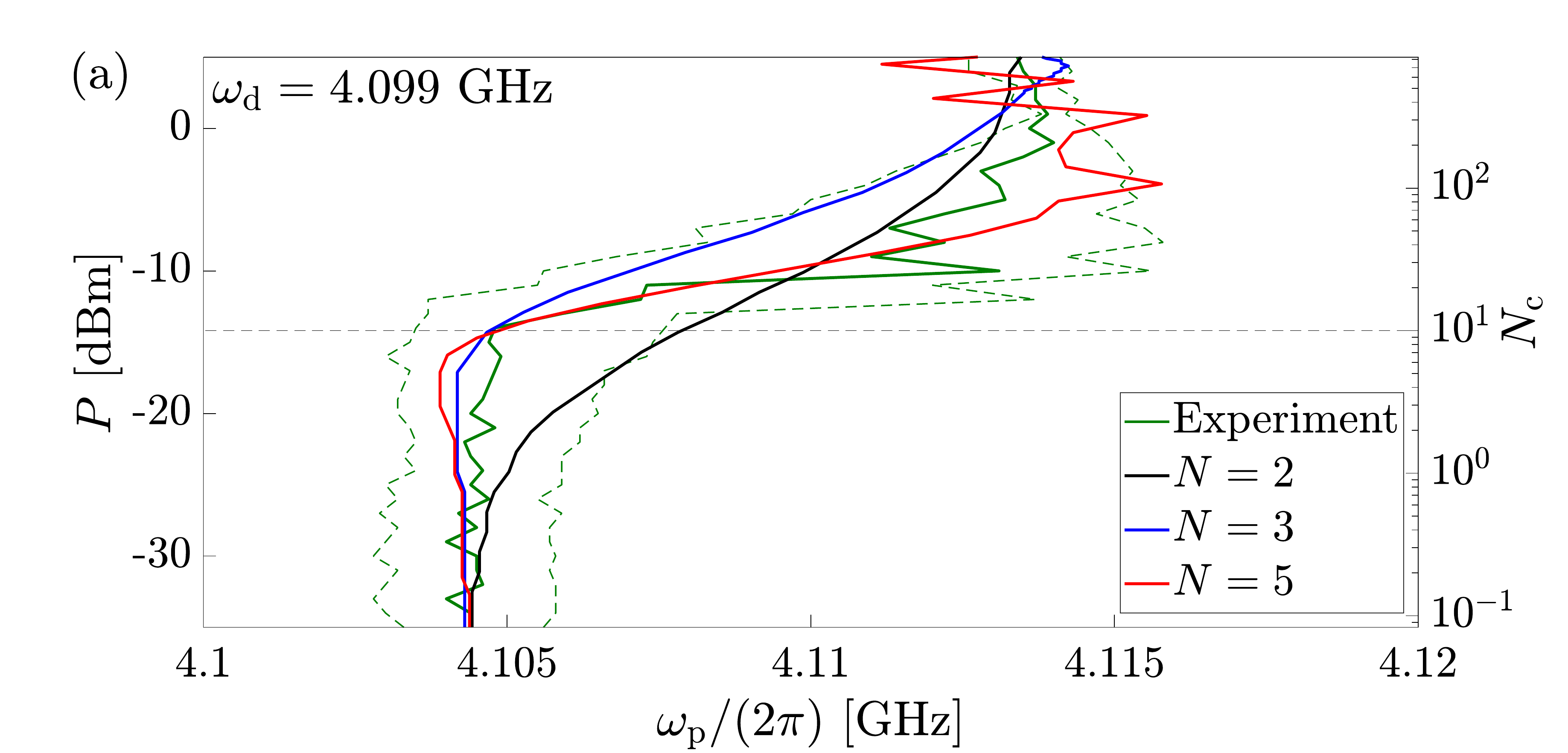}
  \includegraphics[width=1\textwidth]{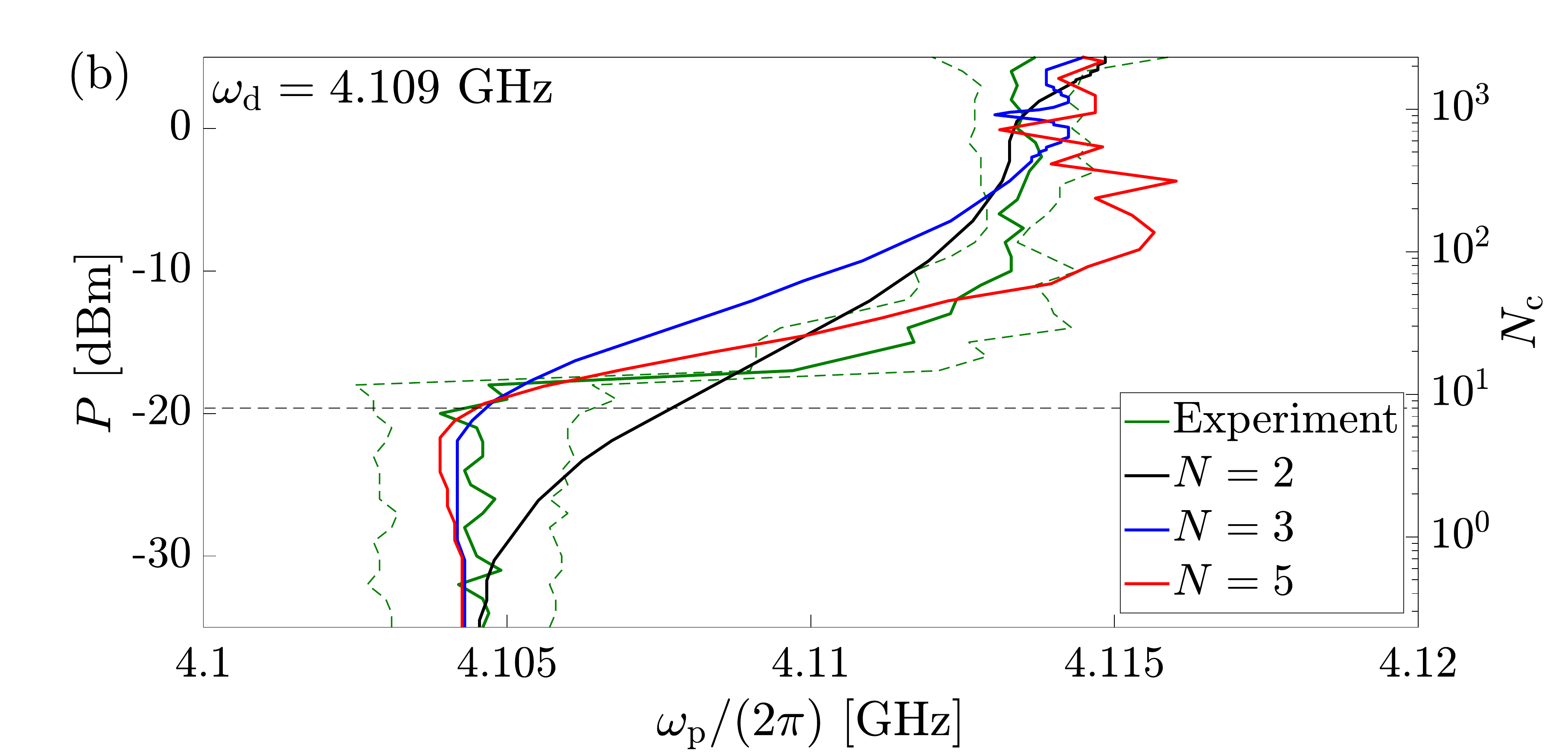}
   \includegraphics[width=1\textwidth]{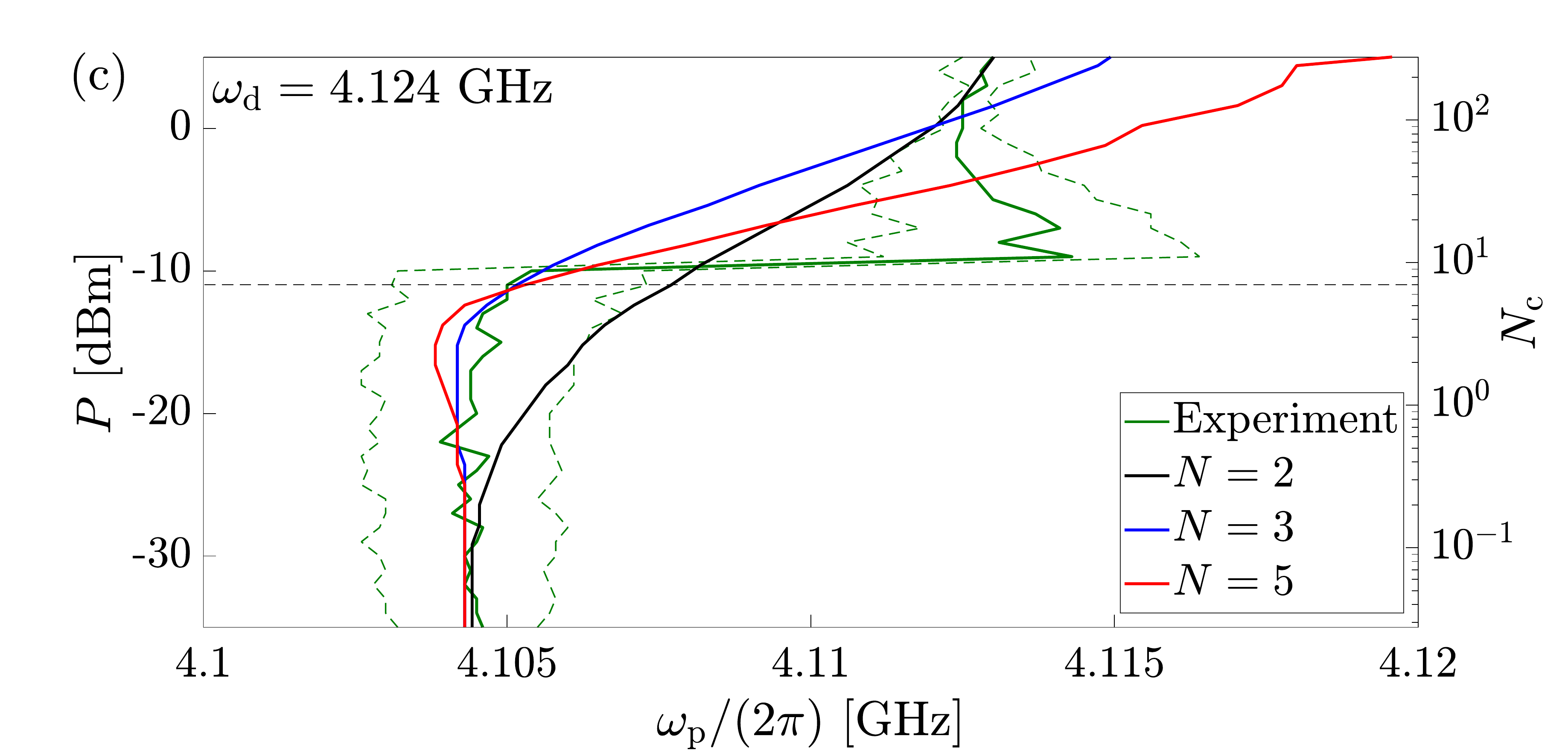}
\caption{The resonance frequency of the system as a function of the probe frequency $\omega_{\rm p}$ and the drive power $P$ (average occupation number $N_{\rm c}$) at three different drive frequencies: (a) $\omega_{\rm d}<\omega_{\rm c}-\chi$; (b) $\omega_{\rm c}-\chi < \omega_{\rm d} < \omega_{\rm c}$; and (c) $\omega_{\rm d}>\omega_{\rm c}$. The green line is the experimental data and the dashed green lines denote the width of the resonance at half maximum. 
We show simulations for three truncations of the transmon. The simulations with $N=5$ have converged up to power value indicated by the black dashed line.}
\label{fig:Experiment4099}
\end{figure}

We study the transition region between the quantum to the classical responses. We are especially interested in the cavity and transmon occupations at the onset of the transition, which we extract from our numerical simulations.
We will show that the transition from quantum to classical is associated with a break-down of the two-level approximation for the transmon.
 The average cavity occupation number is given by $N_{\rm c} = |\alpha_{\rm ss}|^2 +\langle\hat{a}^\dagger\hat{a}\rangle$ where $\hat a$ is here the annihilation operator of the displaced cavity. We have measured the cavity resonance frequency with different drive frequencies. The results are shown in Fig.~\ref{fig:Experiments}. 
We observe that the resonance frequency at low powers is given by $\omega_{\rm c} -\chi$ where $\chi = \chi_0 + \chi_{\rm BS}$ is the vacuum Stark shift and $\chi_{\rm BS} = g^2/(\omega_0+\omega_{\rm c})$ is the vacuum Bloch-Siegert shift, which is a correction imposed by the terms that do not preserve the occupation number. Similar to the simplified two-state model for the transmon, at high powers the cavity resonance occurs at the bare cavity frequency $\omega_{\rm c}$. These is not affected by the drive frequency, as can be seen in Fig.~\ref{fig:Experiments}. We also note that as the detuning $\delta = \omega_{\rm c}-\omega_{\rm d}$ between the cavity and the drive is decreased, the transition starts at a lower power. We refer to the power value at the onset of the transition as the critical power $P_{\rm crit}$, and show it in the inset of Fig.~\ref{fig:Experiments} as a function of the drive frequency. The critical power has a minimum value when $\delta \approx 0$ and increases roughly quadratically as a function of $\delta$.
Even though the critical power depends on the detuning, the transition always occurs close to the same average cavity occupation $N_{\rm c}^{\rm crit} \approx 10$, as shown in Fig. \ref{fig:Experiment4099}. 
This is in accordance with Eq.~(\ref{eq:ass}), if one interprets that the critical cavity occupation obeys the relation $N_c^{\rm crit} \sim P_{\rm crit}(\delta)/\delta^2$. 

When the drive power is increased towards its critical value, the simple two-level approximation for the transmon becomes insufficient and the transmon starts to act as a multilevel quantum system. We demonstrate this by comparing the experimental data with the numerical simulations done with three different truncations for the transmon ($N=2,3$ and 5). In Fig.~\ref{fig:Experiment4099}, 
we show the results in the three relevant different regimes for the drive frequency, $\omega_{\rm d}<\omega_{\rm c}-\chi$, $\omega_{\rm c}-\chi < \omega_{\rm d}<\omega_{\rm c}$, and $\omega_{\rm d}>\omega_{\rm c}$. 
For all values of the drive frequency, we observe that the higher transmon states start to contribute to the resonance frequency as the critical power is approached. This is seen as the emerging deviation between the results obtained with different truncations. However, the correspondence between the simulated and the experimental resonance frequencies improves by increasing the number $N$ of the transmon states included in the simulation. Especially, the simulation with $N=5$ remains within the linewidth of the experimental resonance up to the critical power ($N_{\rm c}^{\rm crit}\sim 10$) for all studied drive frequencies.

In the previous measurement on a different sample we were able to observe the Bloch-Siegert frequency shifts \cite{Pietikainen17}. These shifts appeared in the classical region and had a characteristic oscillatory dependence (nonmonotonic) on the drive power. This dependence is due to the interplay between the counter-rotating drive terms and the coupling terms that do not conserve the occupation number. In the measurements presented here it was not possible to resolve these shifts reliably. The small shifts observed in the frequencies extracted from the spectroscopy data can be attributed to noise and errors in the determination of the minimum of the spectral line, since they appear also at low powers. 

\section{Convergence of the numerical solution}

We study the numerical convergence of the simulations as a function of the truncation number $N$, i.e. the number of transmon states included in our calculations. For each value of $N$, we increase the truncation numbers $N_{\rm c}^{\rm max}$ and $N_{\tau}^{\rm max}$ until further additions do not change the results considerably. However, as the power is increased the addition of a state in the transmon basis has to be accompanied with an inclusion of states also in the other bases. The simulations with $N=5$ transmon states can be continued in this manner only up to power values corresponding to $N_{\rm c} \approx 60$. For larger powers, the required Floquet matrices grow beyond the available memory resources on ordinary tabletop computers. Thus, the corresponding numerical results have not converged, and one should rely on other many-body methods.

\begin{figure}[ht]
  \includegraphics[width=1\textwidth]{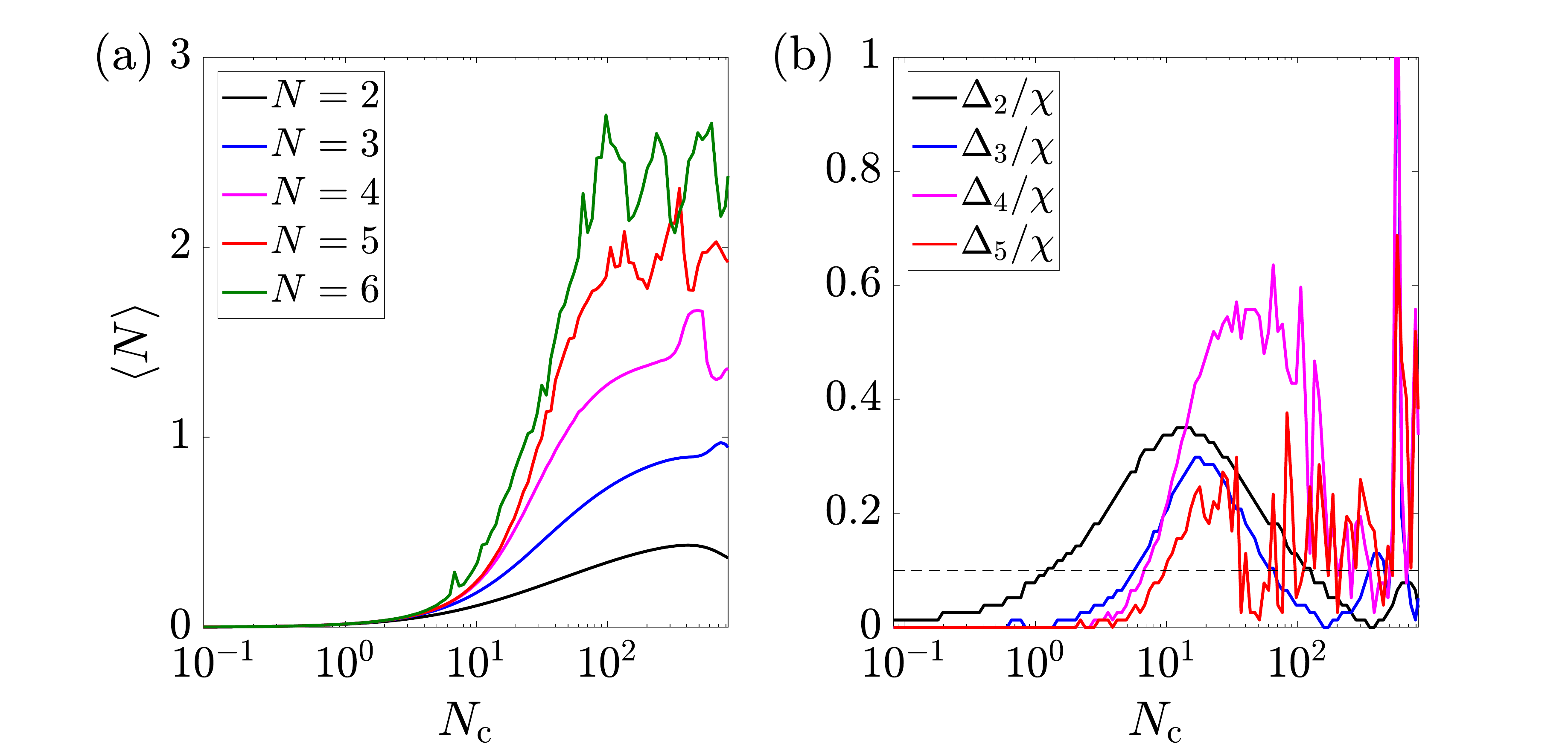}
\caption{Transmon occupation for $\omega_{\rm d}/(2\pi)=4.099$ GHz. (a) The average transmon occupation as a function of the average cavity occupation $N_{\rm c}$. (b) Convergence of the resonance frequency as the number of transmon states included in the simulation is increased. We show the difference $\Delta_N$ between the resonance frequencies obtained with adjacent truncation numbers. The dashed line indicates where the difference is ten percent of the Stark shift.}
\label{fig:AverageN}
\end{figure}

We first calculate the average transmon occupation $\langle N\rangle$. This is shown in Fig.~\ref{fig:AverageN}(a) for different transmon truncation numbers at the drive frequency $\omega_{\rm d}/(2\pi)=4.099$ GHz. We observe that as the drive power approaches the critical value with $N_{\rm c}^{\rm crit}\approx 10$, the numerical solution becomes sensitive to the transmon truncation. 
At the high-power end where $N_{\rm c}\gtrsim 100$, the transmon occupation starts to display oscillatory dependence on the drive power. However, in this region the results have not converged. This is also reflected in the appearance of erratic jumps of the resonance frequency which start around the same value of $N_{\rm c}$ in Fig.~\ref{fig:Experiment4099}.

We also see in Fig.~\ref{fig:Experiment4099} that the experimental quantum-to-classical transition occurs in a very narrow region around $N_{\rm c}^{\rm crit}\sim 10$. For the same power values, the transmon occupation is very sensitive to an increase of power, as depicted in Fig.~\ref{fig:AverageN}(a). We interpret this so that the transition between the quantum and classical responses corresponds to a transition between localized charge states and localized states of the superconducting phase difference across the transmon. A detailed discussion of this will be presented elsewhere~\cite{Pietikainen18}.

We further quantify the convergence of our numerical solution by studying the difference $\Delta_N$ between the numerical resonance frequencies obtained between adjacent truncation numbers of the transmon, i.e. between the solutions with $N+1$ and $N$. Since the observed phenomena occur in the scale of the vacuum Stark shift $\chi$, we express the differences relative to this value. In Fig. \ref{fig:AverageN}(b) we show the relative differences with different truncation numbers of the transmon. We define the simulation as being converged with respect to the transmon truncation when the relative difference is smaller than ten percent. This is depicted in Fig.~\ref{fig:AverageN}(b) with a dashed horizontal line. The power value where the numerical solution crosses this line is shown in Fig.~\ref{fig:Experiment4099} with horizontal lines. For example, when $N=5$ this occurs at $N_{\rm c} \approx 10$, i.e. within the transition region. 

%






\section{Conclusions}
\label{conclusions}

We have studied the cavity and transmon occupations when a multilevel quantum Rabi system is driven across a quantum-to-classical transition. We showed that the transition occurs with the same average cavity occupation independent on the detuning between the drive and the cavity. We also demonstrated that as the drive power approaches its critical value, the transmon escapes the low-energy subspace spanned by its two lowest energy eigenstates. For simulations with five transmon states, we found that the numerical resonance frequency remains within the linewidth from the experimental result up to the powers below the classical regime.


\begin{acknowledgements}
Discussions with D. Golubev and A. Veps\"al\"ainen are gratefully acknowledged. This work was supported the Academy of Finland (Projects No. 263457 and No. 135135), the Finnish Cultural Foundation, Centre of Quantum Engineering at Aalto University (Projects QMET and QMETRO), and the Centres of Excellence LTQ (Project No. 250280), and COMP (Projects No. 251748 and No. 284621). This work used the cryogenic facilities of the Low Temperature Laboratory at OtaNano/Aalto University.
This is a pre-print of an article published in Journal of Low Temperature Physics. The final authenticated version is available online at: \href{https://doi.org/10.1007/s10909-018-1857-8}{https://doi.org/10.1007/s10909-018-1857-8}
\end{acknowledgements}

\bibliographystyle{spphys}       
\bibliography{ConferencePaper}   

\end{document}